# Electromagnetic-field-induced decay of currents in thin-film superconducting rings with photons emission

A.I. Agafonov

Russian Research Center "Kurchatov Institute", Moscow, 123182 Russia

It is shown that currents in thin-film superconducting rings irradiated by coherent microwave fields, can discretely decay with photons emission. These quantized jumps of the supercurrent correspond to destruction one or several magnetic flux quanta trapped in the ring. The ring thickness should be less than the skin-depth of the low-frequency field. The probability of the microwave field-induced single-photon decay of the supercurrent states in these rings is obtained. The angle distributions of the photons emitted by superconducting rings, and the current state lifetimes, depending on the ring sizes and the fluxoid numbers in the initial states, are studied.

## 1. INTRODUCTION

Currents that can exist in multiply connected superconductors in the absence of external magnetic fields, are, in principle, metastable. The current can be varied by the quantum jumps, at which the number of the magnetic-flux quanta, trapped in the superconductor, changes by one or several unites [1,2]. For the spontaneous current decay this quantum jump requires a collective transition of all the Cooper pairs involved in the supercurrents. The probability of such macroscopic fluctuations which cover the whole system with nearly $10^{21}$ particles per cubic centimeter, vanishes that explains the persistence of the superfluid current in the ring.

Besides the macroscopic fluctuation approach [3], the quantized decay of the supercurrent states in the multi-connected structures, in particular, in rings can be due to the radiative decay of these states with the photons emission. However, instability of these current states in the superconducting rings with respect to the electromagnetic vacuum, as far as we know, has not been studied.

The interaction operator of the superconducting current and the electromagnetic field can be represented as:

$$\hat{V} = \int \hat{\mathbf{j}} \hat{\mathbf{A}} d\mathbf{r}, \qquad (1)$$

where $\hat{\mathbf{j}}$ is the operator of the superconducting current density in the ring, $\hat{\mathbf{A}}$ is the operator of the vector potential of the electromagnetic field, and the integration is taken over the ring volume.

Using (1), one can easily verify that for macroscopic values of the mean number of the Cooper pairs $\overline{N} >>> 1$ in the superconductor, the matrix element (1) of the photon decay of the supercurrent in the ring vanishes. This is due to the fact that for the transition $m \to m_1$ between these current states the integral in right-hand side of (1) contains the function $\exp(iN(m - m_1)\varphi)$ with a very large total phase of the coherent condensate. Here $m$ and $m_1$ are the numbers of the magnetic flux quanta in the initial and final states of the condensate, respectively.

In the present paper we show that the perturbation of the superconducting condensate by coherent low-frequency electromagnetic fields can effectively compensate this phase function that leads to finite, really measured lifetimes of the current states in thin-film superconducting rings of relatively small sizes. The probability of the radiation-stimulated single-photon decay of the supercurrent states in these rings is obtained. Numerical results for the angle distributions of the photons emitted by superconducting rings, and the current state lifetimes, depending on the ring sizes and the fluxoid numbers in the initial states, are carried out.

## 2. THE SUPERCURRENT DECAY PROBABILITY

We consider a thin superconducting ring with the inner radius $a >> \lambda$ (where $\lambda$ is penetration depth) and the outer radius $b$ such that $b - a >> \lambda$, and $(b - a)/b << 1$. The thickness $d$ of the ring should be small as compared with the skin depth of the low-frequency field, which is irradiated the ring. Under this condition the field acts simultaneously on the whole condensate. The temperature and magnetic field created by the supercurrent in the ring, are assumed to be small as compared with $T_c$ and the first critical field for the type II superconductors.

Properties of a superconducting ring are completely determined by the condensate wave function $\Psi_m$, where $m$ is the number of the magnetic induction flux quanta (fluxoids), trapped in the ring. So, this function is defined the quantization of the magnetic flux $\Phi_m = m\Phi_0$, the discrete values of the total current $\Phi_m/L$, and the ring energy which is composed of the magnetic energy and the kinetic energy of the condensate, $E_m = E_0 m^2$, where $\Phi_0 = \dfrac{h}{2e}$ is the fluxoid, $L$ is the ring self-inductance and $E_0 = \Phi_0^2/2L$ is the ring characteristic energy.

The superconducting condensate can be described as the superposition of states $|N>$ with different numbers of the Cooper pairs $N$. This superposition has the form of the coherent wave packet [4]:

$$\Psi_m = \sum_N \psi_{Nm}(\boldsymbol{\rho}) c_N |N>,  \qquad (2)$$

where the probability amplitude of the state with $N$ particles is:

$$c_N = \pi^{-1/4} \Delta N^{-1/2} \exp\left(-\dfrac{(N-\overline{N})^2}{2\Delta N^2}\right),  \qquad (3)$$

and the condensate wave function for this state can be written as:

$$\psi_{Nm}(\boldsymbol{\rho}) = \dfrac{1}{\sqrt{\Omega_r}} \exp(iN\phi_m).  \qquad (4)$$

Here $\Omega_r$ is the ring volume, $\phi_m = m\varphi$ is the phase of the single boson wave function, $\varphi$ is the azimuthal angle in cylindrical coordinates tied to the ring ($a \leq \rho \leq b$ and $-d/2 \leq z \leq d/2$).

Note that the coherent wave packet (2)-(4) is normalized, $\int |\Psi_m|^2 d\mathbf{r} = 1$, where the integral is taken over the ring volume.

Using the equations of motion of $N$ and $\phi_m$, the temporal behavior of the phase is defined by [4]:

$$\dfrac{d}{dt} \hbar \nabla \phi_m = \mathbf{F}(t)  \qquad (5)$$

where $\mathbf{F}(t)$ is the total force on the particles. The force induced by the coherent electromagnetic field irradiated the ring, can be written as:

$$\mathbf{F}(t) = 2e\mathbf{A}_0 \cos(\omega_0 t),  \qquad (6)$$

where $\mathbf{A}_0$ is the amplitude of the external field with the frequency $\omega_0$. The latter is assumed to be small as compared with the superconducting gap. Also, we imply that the wavelength of the field is much larger than the characteristic size of the ring, so that the field can be considered as uniform. Without loss of generality, the vector $\mathbf{A}_0$ is considered to be directed against the $y$-axis for the unit vector of which we have $\mathbf{i}_\varphi \mathbf{i}_y = \cos\varphi$.

From (5) and (6) with the initial condition $\phi_m(t=0) = m\varphi$, we find that the phase of the single boson wave function in the initial state of the ring, irradiated the low-frequency field, is:

$$\phi_m(t) = m\varphi - \frac{2eA_0}{\hbar\omega_0}\rho\sin(\varphi)\sin(\omega_0 t). \qquad (7)$$

Of course, this field can change the ring energy in the initial state. To avoid this, we consider the field of low intensities, for which the field correction to the boson velocity $\mathbf{v} = \hbar\nabla\phi_m/m_C$ ($m_C$ is the mass of the Cooper pair) is small. Using (7), we have the restriction on the field amplitude:

$$(\xi_0 b/m)^2 \ll 1, \qquad (8)$$

where

$$\xi_0 = \frac{2eA_0}{\hbar\omega_0}. \qquad (9)$$

Note that, according to (7), on the regular circular motion of the bosons with the velocity $\mathbf{v} = \mathbf{i}_\varphi \hbar m/m_C\rho$ the external field imposes coherent oscillations of the condensate with the very small amplitude which is equal to $(v/\omega_0)(\xi_0 b/m)$.

The transition time between the current states $m \to m_1$ is of the order of $\Delta t \approx \frac{\hbar}{E_m - E_{m_1}} \ll \omega_0^{-1}$. During this time, the low-frequency field does not bring the oscillatory motion of the Cooper pairs in the final state. Therefore, we can consider the transition to the $m_1$ - state of the superconducting condensate, in which the phase of the single boson wave function is:

$$\phi_{m_1} = m_1\varphi. \qquad (10)$$

The off-diagonal matrix element of the current density operator for the superconducting state with $N$ bosons is defined as $\mathbf{j}_{m_1 m}(N) = \psi^*_{Nm_1}\hat{\mathbf{j}}\psi_{Nm}$. Using (7) and (10), and taking into account the normalization of the wave function (4), this matrix element can be written as:

$$\mathbf{j}_{m_1 m}(N) = \frac{e\hbar N}{m_C \Omega_r}\left\{\frac{m+m_1}{\rho}\mathbf{i}_\varphi - \xi_0 \sin(\omega_0 t)\mathbf{i}_y\right\}\exp[iN(m-m_1)\varphi - iN\xi_0\rho\sin(\varphi)\sin(\omega_0 t)] -$$

$$-\frac{1}{4\pi\lambda(N/\Omega_r)^2}\int\frac{d\mathbf{r}_1}{|\mathbf{r}_1 - \mathbf{r}|}\mathbf{j}_{m_1 m}(\mathbf{r}_1, N), \qquad (11)$$

where $\xi_0$ is given by (9), $\lambda$ is the London penetration depth.

The diamagnetic transition current represented by the second term on the right-hand side of (11), is suppressed since the integrand contains the very rapidly oscillating phase function. Using the paramagnetic part of the current transition (11), the single-photon matrix element of the interaction (1) is:

$$V_{m_1 m}^{10}(N) = \frac{e \hbar^{3/2} N}{m_C \Omega_r \sqrt{2\varepsilon_0 \omega_k}} \exp\left[-\frac{it}{\hbar}(E_m - E_{m_1} - \hbar\omega_k)\right] * \frac{2\sin(\frac{kd}{2}\cos(\theta_k))}{k\cos(\theta_k)} (I_1 - I_2), \quad (12)$$

where $\hbar\omega_k$ is the energy of the photon with wave vector $\mathbf{k}$, $\theta_k$ is its polar angle. Taking advantage of $\mathbf{i}_\varphi \mathbf{l}_{\mathbf{k}\sigma} = \sin(\theta_l)\sin(\varphi_l - \varphi)$ and $\mathbf{i}_y \mathbf{l}_{\mathbf{k}\sigma} = \sin(\theta_l)\sin(\varphi_l)$, where $\theta_l$ and $\varphi_l$ are the polar and azimuthal angles of the photon polarization $\mathbf{l}_{\mathbf{k}\sigma}$, Eq. (12) are introduced the notations:

$$I_1(N) = (m + m_1)\sin(\theta_l) \int_a^b d\rho \int_0^{2\pi} d\varphi \sin(\varphi_l - \varphi) *$$
$$\exp\left(-ik_\rho \rho \cos(\varphi - \varphi_k) + iN[(m - m_1)\varphi - \xi_0 \rho \sin(\varphi)\sin(\omega_0 t)]\right), \quad (13)$$

$$I_2 = \xi_0 \sin(\omega_0 t)\sin(\theta_l)\sin(\varphi_l) \int_a^b \rho d\rho \int_0^{2\pi} d\varphi$$
$$\exp\left(-ik_\rho \rho \cos(\varphi - \varphi_k) + iN[(m - m_1)\varphi - \xi_0 \rho \sin(\varphi)\sin(\omega_0 t)]\right). \quad (14)$$

Here $k_\rho = k\sin(\theta_k)$ and $\varphi_k$ is the azimuthal angle of the photon wave vector.

In the weak fields restricted by (8), $I_2 \ll I_1$. Therefore we can constrain the calculation of the integral $I_1$ (13) only. Using the well-known expansion of the plane wave over the Bessel functions, we have:

$$\exp\left(-ik_\rho \rho \cos(\varphi - \varphi_k)\right) = \sum_{p=-\infty}^{+\infty} J_p(k_\rho \rho) \exp(ip(\varphi - \tilde{\varphi}_k)), \quad (15)$$

where $\tilde{\varphi}_k = \varphi_k + \pi/2$, and

$$\exp(-iN\xi_0 \rho \sin(\varphi)\sin(\omega_0 t)) = \sum_{n=-\infty}^{+\infty} J_n(N\xi_0 \rho \sin(\varphi))\exp(-in\omega_0 t). \quad (16)$$

With (15) - (16) the right-hand side of (13) becomes:

$$I_1(N) = \frac{m + m_1}{2i}\sin(\theta_l) \sum_{p=-\infty}^{+\infty} e^{-ip\tilde{\varphi}_k} \int_a^b J_p(k_\rho \rho) d\rho \sum_{n=-\infty}^{+\infty} e^{-in\omega_0 t} \int_0^{2\pi} d\varphi$$
$$J_n(N\xi_0 \rho \sin(\varphi))\left[e^{i(\varphi_l - \varphi)} - e^{-i(\varphi_l - \varphi)}\right]\exp(i[N(m - m_1) + p]\varphi). \quad (17)$$

In (17) it is convenient to pass to the summation only over the positive values of $n$,

$$\sum_{n=-\infty}^{+\infty} e^{-in\omega_0 t} \int_0^{2\pi} d\varphi J_n(N\xi_0\rho\sin(\varphi))\left[e^{i(\varphi_l-\varphi)} - e^{-i(\varphi_l-\varphi)}\right]\exp(i[N(m-m_1)+p]\varphi) =$$

$$\int_0^{2\pi} d\varphi e^{i(N(m-m_1)+p)\varphi} J_0(N\xi_0\rho\sin(\varphi))\left[e^{i(\varphi_l-\varphi)} - e^{-i(\varphi_l-\varphi)}\right] + \quad (18)$$

$$\sum_{n=1}^{+\infty}\left(e^{-in\omega_0 t} + (-1)^n e^{in\omega_0 t}\right)\int_0^{2\pi} d\varphi e^{i(N(m-m_1)+p)\varphi} J_n(N\xi_0\rho\sin(\varphi))\left[e^{i(\varphi_l-\varphi)} - e^{-i(\varphi_l-\varphi)}\right].$$

Introducing $M_\pm = N_p \pm 1$ and $N_p = N(m-m_1) + p$, the integral over the azimuthal angle $\varphi$ in (18) is:

$$I_\varphi = \int_0^{2\pi} d\varphi J_n(N\xi_0\rho\sin(\varphi))\exp(iM_\pm\varphi), \quad (19)$$

that is equal to [5]:

$$I_\varphi = \pi\left\{\left(1+(-1)^n\right)\cos\left(\frac{\pi}{2}M_\pm\right) + i\left(1-(-1)^n\right)\sin\left(\frac{\pi}{2}M_\pm\right)\right\} *$$

$$J_{\frac{n-M_\pm}{2}}\left(\frac{N}{2}\xi_0\rho\right) J_{\frac{n+M_\pm}{2}}\left(\frac{N}{2}\xi_0\rho\right). \quad (20)$$

Eq. (17) with account of (18) and (20) is rewritten as:

$$I_1(N) = -i\pi(m+m_1)\sin(\theta_l)\int_a^b d\rho \sum_{p=-\infty}^{+\infty} e^{-ip\tilde{\varphi}_k} J_p(k_\rho\rho)$$

$$\left\{\sin\left(\frac{\pi}{2}N_p\right)\left[e^{i\varphi_l} J_{-\frac{M_-}{2}}\left(\frac{N}{2}\xi_0\rho\right)J_{\frac{M_-}{2}}\left(\frac{N}{2}\xi_0\rho\right) + e^{-i\varphi_l} J_{-\frac{M_+}{2}}\left(\frac{N}{2}\xi_0\rho\right)J_{\frac{M_+}{2}}\left(\frac{N}{2}\xi_0\rho\right)\right]\right.$$

$$+ 2\sin\left(\frac{\pi}{2}N_p\right)\sum_{k=1}^{\infty}\cos(2k\omega_0 t)\left[e^{i\varphi_l} J_{k-\frac{M_-}{2}} J_{k+\frac{M_-}{2}} + e^{-i\varphi_l} J_{k-\frac{M_+}{2}} J_{k+\frac{M_+}{2}}\left(\frac{N}{2}\xi_0\rho\right)\right] \quad (21)$$

$$\left. - 2\cos\left(\frac{\pi}{2}N_p\right)\sum_{k=0}^{\infty}\sin((2k+1)\omega_0 t)\left[e^{i\varphi_l} J_{k+\frac{1}{2}-\frac{M_-}{2}} J_{k+\frac{1}{2}+\frac{M_-}{2}} + e^{-i\varphi_l} J_{k+\frac{1}{2}-\frac{M_+}{2}} J_{k+\frac{1}{2}+\frac{M_+}{2}}\right]\right\}$$

The braces in the right-hand side of (21) are contained three terms. The first and second terms are nonzero only for odd values $N_p = N(m-m_1) + p$. Hence $M_\pm = N_p \pm 1$ are even, and $\frac{1}{2}(N_p \pm 1)$ are integers. The third term in (21) is nonzero for even $N_p$, and the indices $k + \frac{1}{2} \pm \frac{1}{2} M_\pm$ of the Bessel functions are integers.

Using the properties of the Bessel functions, the first term in the braces of (21) is reduced to the form:

$$e^{i\varphi_l} J^2_{\frac{N(m-m_1)+p-1}{2}}\left(\frac{N}{2}\xi_0\rho\right) - e^{-i\varphi_l} J^2_{\frac{N(m-m_1)+p+1}{2}}\left(\frac{N}{2}\xi_0\rho\right) \qquad (22)$$

In expressions (21) and (22) it is convenient to use the Langer formula for the asymptotic of the Bessel functions with large indices and their arguments [6]:

$$J_\nu(\nu x) = \begin{cases} \dfrac{1}{\pi}\left(\dfrac{arth(s)}{s} - 1\right)^{1/2} K_{1/3}(\zeta), x \leq 1 \\ \dfrac{1}{\sqrt{3}}\left(1 - \dfrac{arctg(s)}{s}\right)^{1/2} (J_{1/3}(\zeta) + J_{-1/3}(\zeta)), x \geq 1. \end{cases} \qquad (23)$$

Here $s = \sqrt{|1-x^2|}$ and

$$\zeta = \begin{cases} \nu(arth(s) - s), x \leq 1 \\ \nu(s - arctg(s)), x \geq 1. \end{cases} \qquad (24)$$

In our case (22) we have:

$$x = \frac{\xi_0 \rho}{m - m_1 + \frac{p \pm 1}{N}}, \quad \nu = \frac{1}{2}(N(m - m_1) + p \pm 1).$$

The most significant result for the decay probability of the current states in the ring with the photon emission can be expected for values $x > 1$. Then, according to (23) - (24), we obtain:

$$J^2_{\frac{N(m-m_1)+p\pm 1}{2}}\left(\frac{N}{2}\xi_0\rho\right) = \frac{4}{\pi N}\left[\xi_0^2\rho^2 - \left(m - m_1 + \frac{p \pm 1}{N}\right)^2\right]^{-1/2} \cos^2\left(\zeta(\rho) - \frac{\pi}{4}\right), \qquad (25)$$

where

$$\zeta = \frac{1}{2}(N(m - m_1) + p \pm 1)\left[\sqrt{x^2 - 1} - arctg\sqrt{x^2 - 1}\right].$$

Replacing the rapidly oscillating function in (25) on its average value, this first contribution to $I_1(N)$ is given by:

$$I_1^{(1)}(N) = -i\frac{2(m+m_1)}{N}\sin(\theta_l)\int_a^b d\rho \sum_{p=-\infty}^{+\infty} e^{-ip\tilde{\varphi}_k} J_p(k_\rho \rho)*$$

$$\left[ e^{i\varphi_l}\left[\xi_0^2\rho^2 - \left(m-m_1+\frac{p-1}{N}\right)^2\right]^{-1/2} - e^{-i\varphi_l}\left[\xi_0^2\rho^2 - \left(m-m_1+\frac{p+1}{N}\right)^2\right]^{-1/2}\right], \quad (26)$$

where $N(m-m_1)+p$ is odd.

Of course, the number of bosons in the superconductor is macroscopically large, and the sum over $p$ in (26) is gained at values $p <<< N$. Then we have:

$$\sum_p{}' e^{-ip\tilde{\varphi}_k} J_p(k_\rho \rho) = \begin{cases} \cos(k\rho\sin(\theta_k)\cos(\varphi_k)), & \text{odd } N(m-m_1) \\ -i\sin(k\rho\sin(\theta_k)\cos(\varphi_k)), & \text{even } N(m-m_1) \end{cases}.$$

Here the prime sign denotes the summation over all even values $p$ for odd values $N(m-m_1)$ and over all odd values $p$ if $N(m-m_1)$ is even. As a result, Eq. (26) is reduced to:

$$I_1^{(1)}(N) = 4\frac{m+m_1}{N}\sin(\theta_l)\sin(\varphi_l)\int_a^b \frac{d\rho}{\sqrt{\xi_0^2\rho^2-(m-m_1)^2}} \begin{cases} \cos(k\rho\sin(\theta_k)\cos(\varphi_k)) \\ -i\sin(k\rho\sin(\theta_k)\cos(\varphi_k)) \end{cases}. \quad (27)$$

If the inner radius of the ring does not differ much from the outer radius, $(b-a)/b << 1$, provided $b-a >> \lambda$, the integral on the right-hand side of (27) gives:

$$I_1^{(1)}(N) = \frac{4(m+m_1)}{kN\sqrt{\xi_0^2 R^2-(m-m_1)^2}} \frac{\sin(\theta_l)\sin(\varphi_l)}{\sin(\theta_k)\cos(\varphi_k)} *$$

$$\begin{cases} \sin(kb\sin(\theta_k)\cos(\varphi_k))-\sin(ka\sin(\theta_k)\cos(\varphi_k)), & \text{odd } N(m-m_1) \\ i(\cos(kb\sin(\theta_k)\cos(\varphi_k))-\cos(ka\sin(\theta_k)\cos(\varphi_k))), & \text{even } N(m-m_1) \end{cases}. \quad (28)$$

We emphasize that the formula (28) can be used only for the transition channels $m \to m_1$, for which $\xi_0 a > (m-m_1)$ и $\xi_0 b > (m-m_1)$. Therefore the value $R$ determined by the mean value theorem for the integral on the right-hand side of (27), is restricted by $a < R < b$.

Now we consider the two remaining contributions to $I_1(N)$, defined by (21). As contrasted with the first term considered above, the Bessel functions entering as pair products in these terms, are independent. At

$$x_{\pm} = \frac{N\xi_0\rho}{N(m-m_1)+p\pm 1\pm 2k} > 1$$

the most slowly varying part of the product $J_{k-\frac{M_-}{2}} J_{k+\frac{M_-}{2}}$ is proportional to the function $\cos(\zeta(x_-)-\zeta(x_+))$, where $\zeta(x_\pm)$ is given by (24). Our calculations show that for the rings with the average number of the Cooper pairs $N > 10^9$ and the particle density of the order of $10^{21} cm^{-3}$ this function is very rapidly oscillating and gives zero mean value. Therefore, for rings of relatively large sizes, say more than a few microns, the contributions of these terms on the right-hand side of (21) are small as compared with the first term (22).

Note that the last statement does not apply to mesoscopic rings with characteristic sizes of the order of the penetration depth.

Taking into account of (8) and (9) and using the condition $x > 1$, we obtain the final restriction on the intensity of the low-frequency field that stimulates the current transition in the ring,:

$$m - m_1 < \frac{2eA_0 R}{\hbar \omega_0} << m. \tag{29}$$

Substituting (28) into (12), the squared modulus of the transition matrix element is reduced to the form:

$$|V_{m_1 m}^{10}(N)|^2 = \frac{2^5 e^2 \hbar^3}{c\varepsilon_0 m_C^2 \Omega_r^2} \frac{(m+m_1)^2}{k^5\left(\xi_0^2 R^2 - (m-m_1)^2\right)} \frac{\sin^2(\frac{1}{2} kd \cos(\theta_k))}{\cos^2(\theta_k)} \frac{\sin^2(\theta_l)\sin^2(\varphi_l)}{\sin^2(\theta_k)\cos^2(\varphi_k)} F, \tag{30}$$

where $c$ is the speed of light in vacuum and

$$F(N) = \begin{cases} [\sin(kb\sin(\theta_k)\cos(\varphi_k)) - \sin(ka\sin(\theta_k)\cos(\varphi_k))]^2, & odd\ N(m-m_1) \\ [\cos(kb\sin(\theta_k)\cos(\varphi_k)) - \cos(ka\sin(\theta_k)\cos(\varphi_k))]^2, & even\ N(m-m_1) \end{cases}. \tag{31}$$

Given (30), we can obtain the probability of the electromagnetic-field-induced single-photon decay of the current state in the thin-film superconducting ring for the transition channel $m \to m_1$:

$$w_{m_1 m} = \frac{2\pi}{\hbar} \sum_N \sum_{\mathbf{k}} |c_{Nm}|^2 <|V_{m_1 m}^{10}(N)|^2>_{pol} \delta(E_m - E_{m_1} - \hbar\omega_k), \tag{32}$$

where $<...>_{pol} = \pi^{-1} \int_0^{2\pi} d\varphi_l ...$ denotes the averaging over the photon polarization.

At first, in (32) we perform the summation over $N$. If the number $m - m_1$ is odd then, using (3) and (31), we have:

$$\sum_N |c_{Nm}|^2 F(N) = 2\sin^2\left(\frac{1}{2}k(b-a)\sin(\theta_k)\cos(\varphi_k)\right). \qquad (33)$$

Accordingly, for the even $m - m_1$ we obtain:

$$\sum_N |c_{Nm}|^2 F(N) = 2\sin^2\left(\frac{1}{2}k(b+a)\sin(\theta_k)\cos(\varphi_k)\right)\sin^2\left(\frac{1}{2}k(b-a)\sin(\theta_k)\cos(\varphi_k)\right). \qquad (34)$$

Now, in the expression (32) the averaging over the photon polarizations can be carried on. Using $\mathbf{k}\mathbf{l}_{\mathbf{k}\sigma} = 0$ and the real polarization vectors $\mathbf{l}_{\mathbf{k}\sigma}$, we obtain:

$$\sin^2(\theta_l) = \left(1 + tg^2(\theta_k)\cos^2(\varphi_l - \varphi_k)\right)^{-1}. \qquad (35)$$

Substituting (35) into (32) with account of (30), the integration over $\varphi_l$ is easily carried out. As a result, the decay probability is reduced to:

$$w_{m_1 m} = 2^{13}\left(\frac{\alpha}{\pi}\right)^4 \left(\frac{2m_e}{m_C}\right)^2 \frac{c\lambdabar_e^2 R_L^3}{\Omega_r^2} \frac{F^0_{m-m_1}(k)}{(m+m_1)(m-m_1)^3\left(\xi_0^2 R^2 - (m-m_1)^2\right)}, \qquad (36)$$

where $\alpha$ is the fine structure constant, $m_e$ is the electron mass, $\lambdabar_e$ is the Compton wavelength of the electron, $R_L = L/\mu_0$, $k = \frac{E_0}{c\hbar}(m^2 - m_1^2)$ is the module of the emitted photon wave vector, and

$$F^0_{m-m_1} = \int_0^{2\pi} d\varphi_k \int_0^{\pi} d\theta_k \sin(\theta_k) F_{m-m_1}(\theta_k, \varphi_k; k). \qquad (37)$$

Here the function $F_{m-m_1}(\theta_k, \varphi_k; k)$ which determines the angular distributions of the emitted photons by the rings, is given by:

$$F_{m-m_1}(\theta_k, \varphi_k; k) = \frac{\sin^2(\frac{1}{2}kd\cos(\theta_k))}{\cos^2(\theta_k)} \frac{\sin^2\left(\frac{1}{2}k(b-a)\sin(\theta_k)\cos(\varphi_k)\right)}{\sin^2(\theta_k)\cos^2(\varphi_k)} *$$

$$\left[|\cos(\theta_k)|\left(1 + \frac{1+\cos^2(\theta_k)}{\sin^2(\theta_k)}\cos(2\varphi_k)\right) - 2\frac{\cos(2\varphi_k)}{tg^2(\theta_k)}\right] * f, \qquad (38)$$

with

$$f = \begin{cases} 1, \text{ odd } m - m_1 \\ \sin^2\left(\frac{1}{2}k(b+a)\sin(\theta_k)\cos(\varphi_k)\right), \text{ even } m - m_1 \end{cases}. \quad (39)$$

The expression for the current decay probability (36)-(39) was obtained for the low-frequency field intensities that satisfy the restriction (29), where $R$ ($a < R < b$) is defined by the mean value theorem for the integral on the right-hand side of (27). That is, this formula can be used only for the transition channels $m \to m_1$ which satisfy the conditions $\xi_0 a > (m - m_1)$ and $\xi_0 b > (m - m_1)$.

In the case $\xi_0 b < (m - m_1)$, the probability of the channel $m \to m_1$ is also finite, but exponentially small in accordance with the asymptotic of the Bessel function $K_{1/3}(\zeta)$ in (23). For this reason, this case does not be interested.

If $\xi_0 a < (m - m_1)$ and $\xi_0 b > (m - m_1)$, then there is the value $\rho$ for which $\xi_0 \rho = (m - m_1)$. Then, in the range $(a, \rho)$ the value $x < 1$, and in the region $(\rho, b)$ $x > 1$, that defines the function (23). The transition matrix element can be found only by numerical methods.

According to (36), the transition probability is inversely proportional to the square of the ring volume. Therefore, for small rings one can expect greater the probability of the superconducting current decay, accompanied by the photon emission.

## 3. ASYMTOTIC OF THE DECAY PROBABILITY

The expressions (36)-(39) can be reduced to a simpler form in the case $k(b-a) \gg 1$. Since the photon wave vector modulus $k = \frac{E_0}{c\hbar}(m^2 - m_1^2)$, for the decay channel $m - m_1 = 1$, that corresponds to the destruction of the one magnetic flux quantum in the final state of the superconducting ring, we obtain:

$$k(b-a) = \frac{\pi(2m-1)}{8\alpha} \frac{a}{R_L} \frac{b-a}{a}.$$

With a relatively large number of the magnetic flux quanta trapped in the ring in the initial state, this case can be realized that allows to make the replacement:

$$\frac{\sin^2\left(\frac{1}{2}k(b-a)\sin(\theta_k)\cos(\varphi_k)\right)}{\sin^2(\theta_k)\cos^2(\varphi_k)} \Rightarrow \frac{\pi}{2}k(b-a)\delta(\sin(\theta_k)\cos(\varphi_k)). \quad (40)$$

Substituting (40) into (38) - (39), we obtain:

$$\sin(\theta_k) F_{m-m_1}(\theta_k, \varphi_k; k) = \frac{\pi}{2} k(b-a) G_{m-m_1}(\theta_k, k) \left[ \delta(\varphi_k - \frac{\pi}{2}) + \delta(\varphi_k - \frac{3\pi}{2}) \right], \quad (41)$$

where the function $G_{m-m_1}$ determined the polar angle distribution of the photons emitted by the superconducting rings, is:

$$G_{m-m_1}(\theta_k, k) = \frac{1 - \cos(kd \cos(\theta_k))}{1 + |\cos(\theta_k)|} \begin{cases} 1, & \text{odd } m - m_1 \\ \frac{1}{2}, & \text{even } m - m_1 \end{cases}. \quad (42)$$

It is easy to verify that the two $\delta$–functions, $\delta(\theta_k)$ and $\delta(\theta_k - \pi)$, which are also contained in the right-hand side of (40), do not contribute to the angular distribution $G_{m-m_1}(\theta_k)$.

Taking into account (40) - (42), the decay probability (36) for the rectangular cross section ring is reduced to:

$$w_{m_1 m} = 2^{10} \frac{\alpha^3}{\pi^4} \left( \frac{2m_e}{m_C} \right)^2 \frac{c \lambdabar_e^2 R_L^2}{d^2 (b-a)(b+a)^2} \frac{G^0_{m-m_1}}{(m-m_1)^2 \left( \xi_0^2 R^2 - (m-m_1)^2 \right)}, \quad (43)$$

where $G^0_{m-m_1}(k) = \int_0^\pi G_{m-m_1}(\theta_k, k) d\theta_k$.

Note that the probability (43) was obtained at $k(b-a) \gg 1$. Therefore, this expression can be called the asymptotic one.

## 4. LIFETIME CALCULATIONS AND DISCUSSION

Below we use the results for the self-inductance $R_L = L/\mu_0$ of superconducting thin flat rings as a function of the quantity $a/b$, shown in Fig. 2 of the work[7]. The parameter of the curves presented, is $\Lambda/b$, where $\Lambda = \lambda^2 / d$ is the two-dimensional effective penetration depth and the ring thickness $d < \lambda/2$. In our calculations $d = 0.4\lambda$ is used. Then the parameter of these curves is reduced to $2.5\lambda/b$, where $b$ is the outer radius of the ring. As the penetration depth, a typical value $\lambda = 2*10^3 A^0$ for type-II superconductors is used. Accordingly, the thin-film ring thickness is equal to $d = 800 A^0$. For all rings studied below, their inner radii $a = 0.8b$, keeping in mind that $b - a \gg \lambda$.

The decay probabilities are calculated for the rings with four different outer radii. Introducing the notation $R_L = \beta a$ and using the data [7], we obtain $\beta = 7.1$ for the ring with $b = 5\mu$, $\beta = 4.6$ with $b = 16.7\mu$, $\beta = 3.7$ at $b = 50\mu$ and $\beta = 3.4$ with $b = 500\mu$.

Assuming $m_C = 2m_e$, the decay probability (36) for the rings considered, is rewritten as:

$$w_{m_1 m} = K_0 \frac{\beta^3}{b} \frac{F^0_{m-m_1}(k)}{(m+m_1)(m-m_1)^3 \left(\xi_0^2 R^2 - (m-m_1)^2\right)}, \qquad (44)$$

where $K_0 = 6.673*10^{-10} \frac{m}{s}$.

Similarly, the asymptotic expression (43) can be presented as:

$$w_{m_1 m} = K \frac{\beta^2}{b} \frac{G^0_{m-m_1}(k)}{(m-m_1)^2 \left(\xi_0^2 R^2 - (m-m_1)^2\right)}, \qquad (45)$$

with $K = 2.821*10^{-8} \frac{m}{s}$.

The probability vanishes in absence of the low-frequency field, as it follows from (7)-(8), (44) and (45). Therefore it is important to estimate the field intensity. Let $\xi_0 R = 3.5$, that is the three current transition channels $m - m_1 = 1, 2, 3$ are permitted with the destruction of the one, two and three magnetic flux quanta in the final state of the superconducting ring, respectively. From (8) we have $A_0 = 1.8\hbar\omega_0 / eR$. For the field frequency $\omega_0 = 10^{12} s^{-1}$ and $R = 10\mu$ we obtain the coherent field amplitude $A_0 \approx 10^2 V/m$ that corresponds to the intensity $\approx 1.3 mW/cm^2$.

Fig. 1 shows the polar angle distribution of the photons emitted by the superconducting rings with the outer radius $b = 5\mu$ for the decay channel $m - m_1 = 1$, which corresponds to the destruction of the one magnetic flux quantum in the final state. The microwave field amplitude corresponds to the value $\xi_0 R = 1.3$. The distribution is always symmetric with respect to the angle $\theta_k = \pi/2$, at which photon emission is absent. The ring inductance is equal to $L = 7.1\mu_0 a$, and the energy of the photon emitted by the ring is $\hbar\omega_k = E_0(2m-1)$, where $E_0 = \Phi_0^2/2L$. At $m = 61$ this energy is equal to $\hbar\omega_k = 45.25 eV$. Oscillations in the distribution is gained with the increasing the fluxoid number in the initial state.

The azimuthal angle distribution of the emitted photons represents always two very narrow peaks, whose maxima occur at the angles $\varphi_k = \pi/2$ and $\varphi_k = 3\pi/2$. These values correspond to the $y$-axis along which the low-frequency field was directed. That is, this distribution is determined by the vector $\mathbf{A}_0(t)$, and the wave vectors of the emitted photons lie in the $(yz)$-plane. With the increasing the fluxoid number in the initial state, these peaks become more narrow, passing into the distribution (41).

The above features in the angular distributions have been found for all the rings investigated.

The lifetime of the $m$ supercurrent state in ring with respect to the decay channel $m \to m_1$ is:

$$\tau_{m_1 m} = w^{-1}_{m_1, m}.$$

Fig. 2 demonstrates the fluxoid number dependence of the lifetime of the transition $m \to m-1$ for these four different rings. These calculations were performed with the use of the formula (44). Note a good agreement of these data with the results obtained from the asymptotic expression (45). The general feature is that with the increasing of the fluxoid number at first $\tau_{m-1,m}$ decreases sharply and then oscillates about some mean value which is typical for each ring. The amplitude of these oscillations decreases, though weakly, with the increasing of $m$. Using (45), this allows to determine the asymptotic lifetime of the current state,

$$\tilde{\tau}_{m_1 m} = K^{-1} \frac{b}{\beta^2} (m - m_1)^2 \left( \xi_0^2 R^2 - (m - m_1)^2 \right) \begin{cases} \frac{1}{2}, & \text{odd } m - m_1 \\ 1, & \text{even } m - m_1 \end{cases}. \qquad (46)$$

Thus, $\tilde{\tau}_{m-1,m} = 1.25 s$ for the ring with the outer radius $b = 5\mu$, whereas $\tilde{\tau}_{m-1,m} = 680 s$ at $b = 500\mu$.

With the increasing of the low-frequency field intensity several current decay channels $m \to m_1$ can be permitted with the destruction of $m - m_1$ fluxoids in the final states of the superconducting ring. Then the total probability of the superconducting current decay is given by:

$$w_m = \sum_{m_1} w_{m_1, m}. \qquad (47)$$

For large $m$ each ring can be completely characterized by the asymptotic lifetime (46), as follows from Fig. 2. Table I shows the values $\tilde{\tau}_{m_1, m}$ for the allowed current transitions depending on the microwave field amplitude, which is given in units $\xi_0 R$. At low intensities $\xi_0 b < 1$ and, respectively, $\xi_0 R < 1$, current transitions are not permitted. Hence the superconducting current in the ring is persistent.

If $\xi_0 R > 1$ and $\xi_0 b < 2$, the only current transition $m \to m-1$ is permitted, as shown in Table I. We cannot calculate the minimum lifetime of this transition because the mean value theorem for the integral on the right-hand side of (27) was used. When $\xi_0 R > 2$ and $\xi_0 b < 3$, two current transitions with $m \to m-1$ and $m \to m-2$ can be observed. At $\xi_0 R = 2.3$ the lifetime of the first transition is 7.76 s, and for the second one $\tau_{m_1,m}^{as}$ = 18.67 s. With the increasing of the low-frequency field intensity the total decay probability of the superconducting current in the ring decreases. This clearly demonstrates the third and fourth rows of the Table I, which represent the asymptotic lifetimes at the field intensities that allow the three and four current transitions, respectively.

## 5. SUMMARY

The found effect of the microwave field-induced single-photon decay of the supercurrent states can be observed only in the thin-film rings, the thickness of which is less than the skin-depth of this low-frequency field. Under this condition the coherent field can cause the collective transition of all the Cooper pairs involved in the supercurrents. The latter decays by quantum jumps that correspond to destruction of one or several fluxoids trapped in the ring. Of course, in massive rings the effect cannot occur.

It was obtained that the decay channel $m \to m-1$ gives the main contribution to the lifetime of the current state. With the decreasing of the ring sizes the lifetime of this channel falls off, as shown in Fig. 2. Hence, for experimental observations of the current decay the small rings should be used. In this regard, the mesoscopic rings are of interest. In addition, with the decreasing of the ring sizes the ring inductance decreases, and, respectively, the characteristic energy, $E_0 = \Phi_0^2 / 2L$, increases. The energy of the emitted photon, $\hbar \omega_k = E_0 (2m-1)$, can be large, of the order of $10^2$ eV.

The field intensity dependence of the total probability of the superconducting current decay is the narrow peak, maximum of which falls on the field amplitude $\xi_0 R \approx 1$. Therefore, fine-tuning of the radiation intensity stimulated the current transitions, is required for each ring.

Table I: The asymptotic lifetimes of the superconducting current transitions for the ring with the outer radius $b = 5\mu$.

| $\xi_0 R$ | $\tilde{\tau}_{m-1,m}$ (s) | $\tilde{\tau}_{m-2,m}$ (s) | $\tilde{\tau}_{m-3,m}$ (s) | $\tilde{\tau}_{m-4,m}$ (s) |
|---|---|---|---|---|
| 1.3 | 1.25 | - | - | - |
| 2.3 | 7.76 | 18.67 | - | - |
| 3.3 | 17.89 | 99.71 | 30.77 | - |
| 4.3 | 31.64 | 209.70 | 154.50 | 144.14 |

**Figure captions:**

Fig. 1. The polar angle distribution of the photons emitted by the superconducting rings with the outer radius $b = 5\mu$, for the decay channel $m \rightarrow m-1$ and the three fluxoid numbers in the initial states. The parameter $\xi_0 R = 1.3$.

Fig 2. The fluxoid number dependence of the lifetime of the transition $m \rightarrow m-1$ for the four rings with different outer radii. The parameter $\xi_0 R$ is the same as in Fig.1.

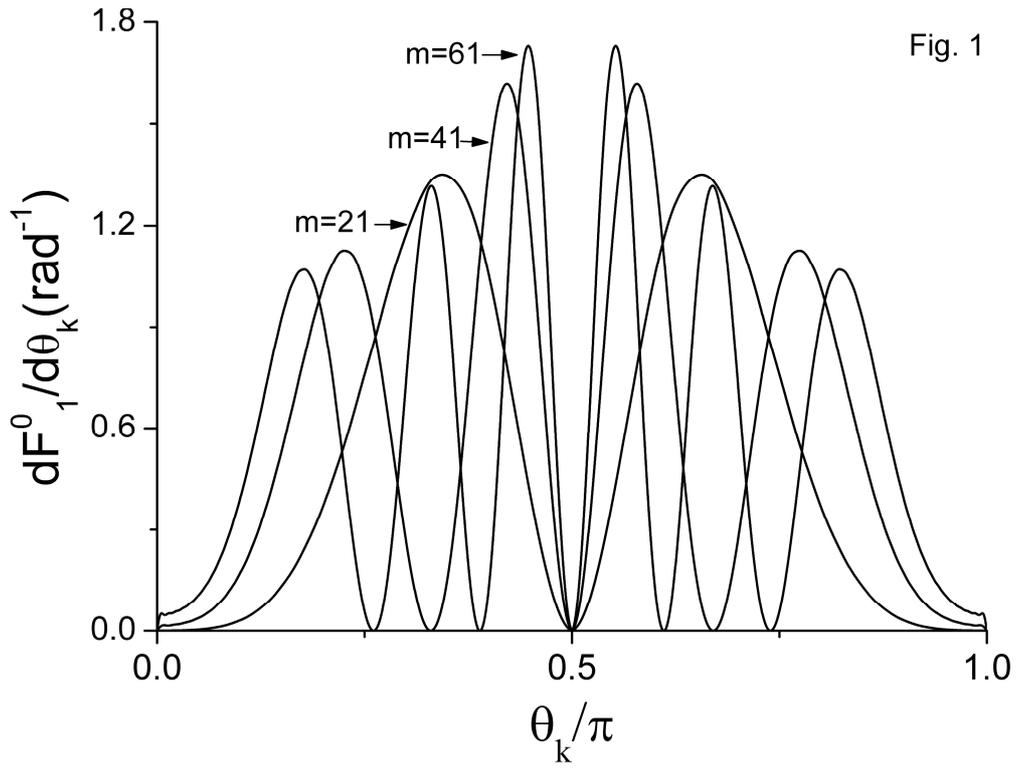

Fig. 1

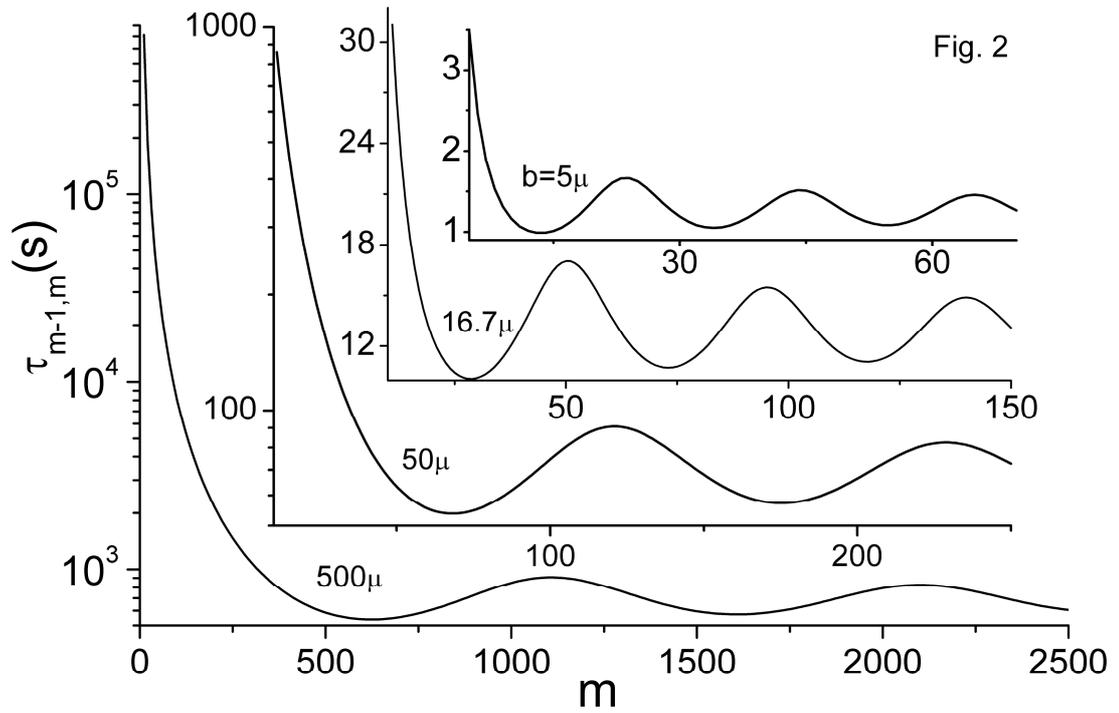

Fig. 2